# Title: Dynamic Computerized Tumbling-E Testing for Temporal Reliability of Human Sequential Perceptual Decisions

Authors: Avneek Sandhu[1,2], and Bin Hu[1,3]

**Affiliations**:

1. Canadian Open Digital Health (OpenDH) program, University of Calgary, Calgary, Alberta, Canada T2N 4N1
2. Faculty of Science, University of Calgary, Calgary, AB T2N 4N1, Canada
3. Division of Translational Neuroscience, Department of Clinical Neurosciences, Hotchkiss Brain Institute, University of Calgary, Calgary, AB T2N 4N1, Canada

**To whom corresponding should be addressed:**
**Professor Bin Hu MD. Ph.D.**
**Suter Professor for Parkinson's Disease Research**
**Founder and Director**
**Open Digital Health (OpenDH) Program of University of Calgary**
**Email:** hub@ucalgary.ca

Abstract:

OBJECTIVES: Visual acuity and tumbling-E tasks are often treated as static threshold measures, yet sequential perceptual decisions unfold over time. A computerized tumbling-E task preserves response latency, timeouts, and stimulus-size adaptation, creating a temporal reliability dataset rather than only a chart-line score. This matters for human-AI comparison because the Temporal Hallucination Index (THI) shows how static accuracy can obscure delays, drift, persistence, and unstable convergence.

METHODS: We curated trial-level human data from a computerized dynamic tumbling-E task. On each trial, a single E optotype appeared in one of four orientations, participants selected the perceived direction or timed out, and stimulus size was automatically adjusted through an adaptive staircase. Primary outcomes were reaction time, timeout rate, delay rate above a 3-second budget, and observable THI based on delay and timeout components.

RESULTS: The final dataset included 1,154 valid trials from 21 human identifiers across 77 sessions. There were 1,078 non-timeout responses and 76 timeouts, giving a 6.6% timeout rate. Non-timeout reaction times centered near 1.5 seconds (mean 1546 ms; median 1506 ms; IQR 1306-1713 ms), with only 3 responses exceeding 3,000 ms. Adaptation was dominated by smaller-next-stimulus transitions (89.2%). Mean arcminutes declined from 29.42 at trial 0 to 5.04 at trial 19, supporting convergence near a 20/20-level optotype without clinical acuity diagnosis.

CONCLUSIONS: This dataset converts a tumbling-E visual task into a temporally resolved human perceptual-decision benchmark. Its novel contribution is automatic capture of staircase behavior, response timing, timeouts, and trial-level reliability signals. The human data show fast timing and smooth adaptation toward threshold, establishing a human-only baseline for future comparison with artificial agents.

## Introduction

Visual acuity testing has historically relied on optotypes that decrease in size until a participant can no longer identify them reliably. Snellen-style scoring relates optotype size to viewing distance, while logMAR and Early Treatment Diabetic Retinopathy Study (ETDRS) style charts improved research standardization by implementing regular size steps, standardized spacing, and rigorous scoring conventions [1-6]. Tumbling-E tasks are particularly valuable when the goal is to mitigate language, literacy, and letter-recognition biases [2, 7, 8]. Rather than naming a character, the participant identifies the orientation of a rotated E, making it a pure directional recognition paradigm [8, 9].

The present computerized implementation differs from traditional static charts in three fundamental ways. First, it is entirely digital: a single E-shaped optotype is displayed on a high-resolution display, replacing the printed chart rows read under manual examiner control [10, 11]. Second, it is dynamic: stimulus size changes automatically during the sequence, allowing the task to approach a perceptual threshold through an adaptive staircase procedure rooted in classical psychophysics [12-15]. Third, it is temporally instrumented: every individual trial stores a rich array of telemetry, including timestamps, user session indices, trial sequence numbers, orientation choices, exact reaction times, timeout states, pixel sizes, and visual angle arcminute values [16]. These features render the dataset highly suitable for temporal reliability analysis, extending utility far beyond simple, static visual-threshold estimation [16, 17].

The necessity for temporal reliability metrics is directly motivated by the Temporal Hallucination Index (THI) framework [18]. This paradigm posits that both human perception and artificial intelligence (AI) responses unfold dynamically over time, and that static performance metrics can fail to capture critical behavioral instabilities such as processing drift, flip-flopping, response persistence, and sudden timeouts [18-21]. In cognitive science and computer vision, sequential sampling models—such as the drift-diffusion model (DDM)—demonstrate that speed and accuracy are inextricably linked through a continuous evidence-accumulation process [22-25]. Static metrics hide these underlying dynamics. Consequently, high-quality human behavioral data serve as an essential reference distribution, illustrating what stable sequential perceptual decisions look like as visual task difficulty progresses toward the biological limit [25-27].

This article extracts the human subset from the exported results table and presents it as a publication-ready dataset manuscript. The scientific inquiry is not whether this system replaces a clinical visual-acuity examination, but whether a computerized dynamic tumbling-E task can generate a reliable human temporal baseline for future human-AI perceptual decision studies [28-30].

# Methods

## Experimental Design

This study was a secondary dataset analysis of an exported results table from a computerized visual task. The experimental unit was the individual computerized tumbling-E trial. Data were collected from human subjects who provided written informed consent prior to participation. The study protocol was reviewed and approved by the Institutional Review Board (IRB) of the University of Calgary. All research was conducted in accordance with the ethical standards laid down in the 1964 Declaration of Helsinki and its later amendments.

The workflow was designed to align with a datasets/benchmarks/protocols manuscript framework rather than a clinical diagnostic validation paper. Accordingly, the report focuses on data extraction, operational definitions, reproducible filtering, temporal behavior, and task interpretation. The computerized task should be conceptualized as a research benchmark for sequential perceptual reliability, not as a substitute for a calibrated clinical ETDRS/logMAR examination.

## Computerized dynamic tumbling-E task

On each trial, the participant viewed a single E optotype on a screen, rotated randomly to one of four orientations: up, right, down, or left. The participant selected the perceived direction, or the trial ended without a valid response. Unlike a traditional chart, the system does not depend on a fixed printed row or a manual examiner decision for each step. Instead, it automatically adjusts the E size over the course of the session based on psychophysical principles, creating an adaptation curve [12, 14]. Early trials begin with larger, highly resolvable optotypes; subsequent trials progressively become smaller and more difficult as the staircase approaches the participant's perceptual threshold [13, 15].

The approximate viewing distance for this dataset was 30 cm. This distance is critical because visual angle is a mathematical function of physical optotype size and distance from the eye [3, 28]. The exported data included an Arcmin column, which was utilized directly for visual-angle analysis rather than recalculating arcminutes from pixel size, as device-specific screen dimensions and pixel density were unavailable in the source table. The viewing distance is therefore reported as approximate, and all 20/20 interpretations are treated as task-equivalent rather than clinical diagnostic measures [5, 29].

Reaction time (RT) was defined as the elapsed time, in milliseconds, between stimulus presentation and the participant's recorded orientation response, a parameter critical to measuring cognitive and perceptual workloads [23, 24]. A timeout was recorded when no valid orientation choice was made within the response budget. The experimental design utilized a 3-second time constraint for humans; therefore, a delay was operationalized as a numeric reaction time exceeding 3,000 ms [22, 26].

**Data Collection and Analysis**

To generate a clean, human-only analytic dataset, records were excluded when the User ID exactly matched known test, AI, or example identifiers (e.g., *Example_1, g001, G1505, TEST, GPT_01, G2547, C6084, GPT, C8480, C1505, AI test, CMC_test*). All other records were retained for human-only analysis. Exact duplicate exported rows were removed. The final analytic dataset included only records with interpretable trial, response, stimulus-size, and arcminute fields.

The retained dataset included 1,154 valid trials from 21 human identifiers across 77 sessions, spanning 2025-10-21 15:32:15 to 2025-10-22 13:20:31. Retained human identifiers were: 64, 368, 530, 688, 731, 842, 951, 1066, 1843, 4500, 5465, 5513, 5689, 6480, 7364, 7841, 7895, 8306, 9146, 9443, TY2939.

**Column dictionary**

- **Timestamp:** Date and time at which the trial event was recorded; used to preserve temporal ordering.
- **User ID:** Pseudonymous identifier used for filtering and aggregation; excluded IDs were removed exactly.
- **Session ID:** Within-user session label; used to summarize repeated sequential runs.
- **Trial:** Ordinal position of the trial within the session; used for adaptation-curve analysis.
- **Choice:** Participant response: up, right, down, left, or none. None indicates no recorded orientation response.
- **Correct:** Exported correctness flag. Non-informative for accuracy analysis in this dataset; the manuscript emphasizes timing, timeout, stimulus size, and arcminute outcomes.
- **RT (ms):** Reaction time in milliseconds for non-timeout responses, or timeout when no response was recorded.
- **Stimulus Size (px):** Screen-pixel size of the E stimulus; used as one difficulty scale.
- **Arcmin:** Exported visual-angle value in arcminutes; interpreted as the primary size/difficulty scale.

- **Derived variables:** Timeout, numeric RT, delay over 3,000 ms, within-session transition direction, and observable THI components.

**Outcome definitions**

The primary descriptive outcomes were: (1) non-timeout reaction time, (2) timeout rate, (3) delay rate above the 3-second human time budget, (4) stimulus-size and arcminute trajectories across trial number, and (5) within-session transition direction. A smaller-next transition indicated that the subsequent trial in the same session had a smaller stimulus size, consistent with increasing difficulty. A larger-next transition indicated a larger subsequent stimulus size, consistent with a staircase correction or recovery step [12, 13].

The observable THI was calculated from directly measurable human instability events as:

*Observable THI = (Delay rate + Timeout rate) / 2*

This narrow index was utilized because the full THI formulation includes drift and persistence, which require reliable correctness labels or a task-specific ground-truth key [18, 19]. The exported Correct column was unsuitable for accuracy-based inference; therefore, drift and persistence were excluded from inferential outcomes in this human-only analysis.

**Visual-angle and 20/20 interpretation**

The 20/20 convention corresponds to resolving a critical detail of approximately 1 minute of arc. In standard optotype geometry, the total optotype height is exactly five times the critical detail (the stroke width and gaps), meaning a 20/20 optotype subtends a total of 5 arcminutes of visual angle [1, 3, 5]. If the exported Arcmin value represents total optotype height, a 5-arcminute E approximates a 20/20-level stimulus. Under this assumption, a mean arcminute value of 5.04 at trial 19 equates to approximately 20/20.2.

This supports the conclusion that the adaptation curve converged near a 20/20-level stimulus size. This should not be reported as measured clinical acuity, as the dataset was collected on a computerized near-screen task at approximately 30 cm without fully documented chart luminance, screen calibration, refraction, or monocular/binocular viewing conditions [6, 11, 30].

**Statistical analysis**

Analyses were descriptive. Counts and percentages were reported for filtering outcomes, response choices, timeouts, and transition categories. Reaction time was summarized using mean, standard deviation, median, and interquartile range (IQR) among non-timeout numeric responses. Stimulus size and arcminute values were summarized overall and by trial number. Snellen-equivalent interpretations were derived solely as approximate task-equivalent values using the assumption that total optotype height in arcminutes divided by five approximates the minimum angle of resolution (MAR) [3, 4].

## Results

### Human-only dataset construction

After applying the exact exclusion list and removing duplicate exported rows, the human-only analytic subset contained 1,154 valid trials, 21 retained human identifiers, and 77 sessions. The retained identifiers were: 64, 368, 530, 688, 731, 842, 951, 1066, 1843, 4500, 5465, 5513, 5689, 6480, 7364, 7841, 7895, 8306, 9146, 9443, TY2939.

Figure 1. Human-only DDAY filtering workflow

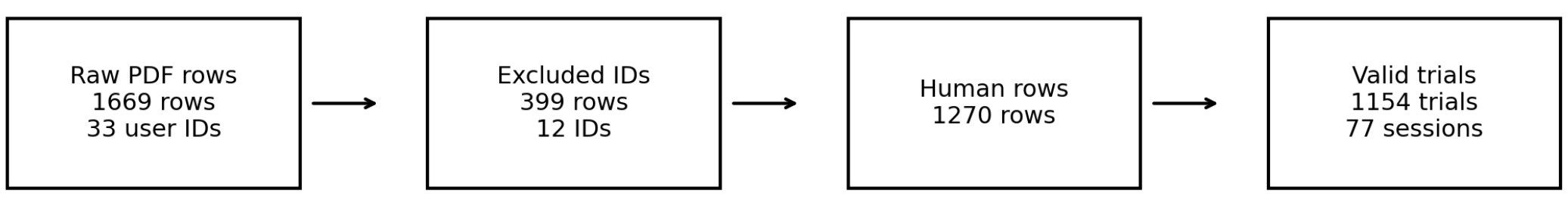

Exact duplicate rows removed: 109

Excluded by exact User ID: Example_1, g001, G1505, TEST, GPT_01, G2547, C6084, GPT, C8480, C1505, AI test, CMC_test

***Figure 1. Human-only data filtering workflow.*** *The exported table was filtered by exact User ID, de-duplicated, and reduced to valid human trials with interpretable trial, response, stimulus-size, and arcminute fields.*

## Human response latency and timeout behavior

The human subset contained 1,078 non-timeout responses and 76 timeouts. The timeout rate was 6.6%. Non-timeout reaction times were centered near 1.5 seconds: mean 1546 ms (SD 350 ms), median 1506 ms, and IQR 1306 to 1713 ms. Only 3 responses exceeded the 3-second human time budget, corresponding to 0.28% of non-timeout responses.

This reaction-time distribution supports the interpretation that most retained human responses were deliberate, rapid perceptual decisions rather than delayed or unstable responses [23]. In a dynamic staircase task, RT is not merely a nuisance variable; it is a critical temporal marker indicating whether participants can maintain decision speed and structural focus as visual difficulty increases [22, 25].

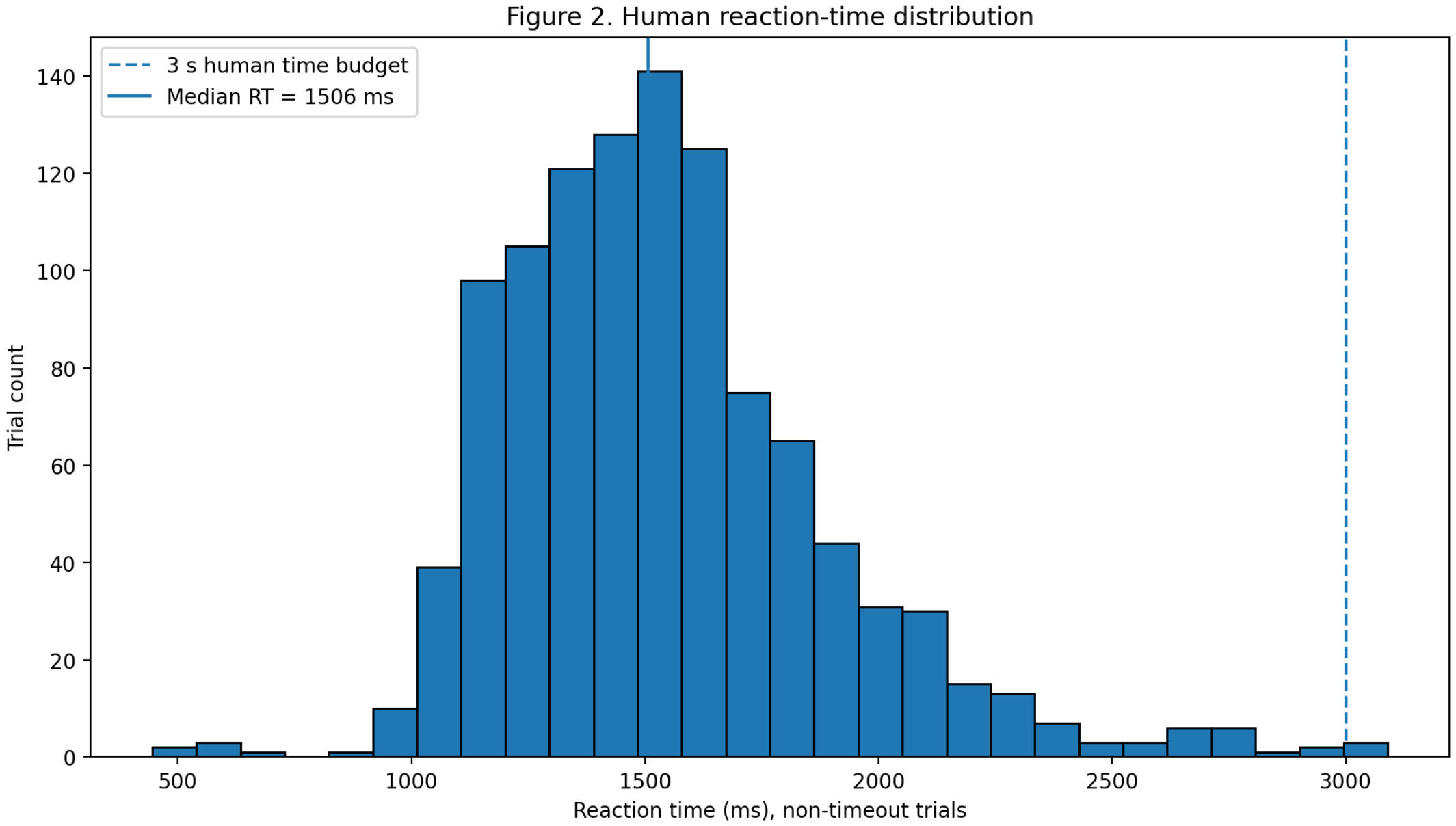

***Figure 2. Human reaction-time distribution.*** *Non-timeout human responses clustered around approximately 1.5 seconds, with very few responses beyond the 3-second human time budget.*

**Automatic staircase convergence and adaptation curve**

The adaptation curve showed progressive movement toward smaller optotypes. Median retained stimulus size was 20 px (range, 7 to 63), and median arcminute value was 11.52 (range, 3.40 to 32.02). Across 1,077 within-session transitions, 89.2% moved to a smaller next stimulus, 10.2% were unchanged, and 0.6% moved to a larger next stimulus.

The mean arcminute value decreased from 29.42 at trial 0 to 5.04 at trial 19. This pattern is the expected shape of an adaptive perceptual staircase: an initial rapid descent through easier, highly visible levels followed by a flatter terminal region near threshold [12, 14]. Under the standard optotype-height assumption, the trial 17–19 mean of 5.07 arcminutes corresponds approximately to 20/20.3, and trial 19 alone corresponds approximately to 20/20.2. Thus, the data are consistent with convergence near a 20/20-level stimulus size, functioning as a computerized task-equivalent interpretation rather than a clinical diagnosis [5, 10].

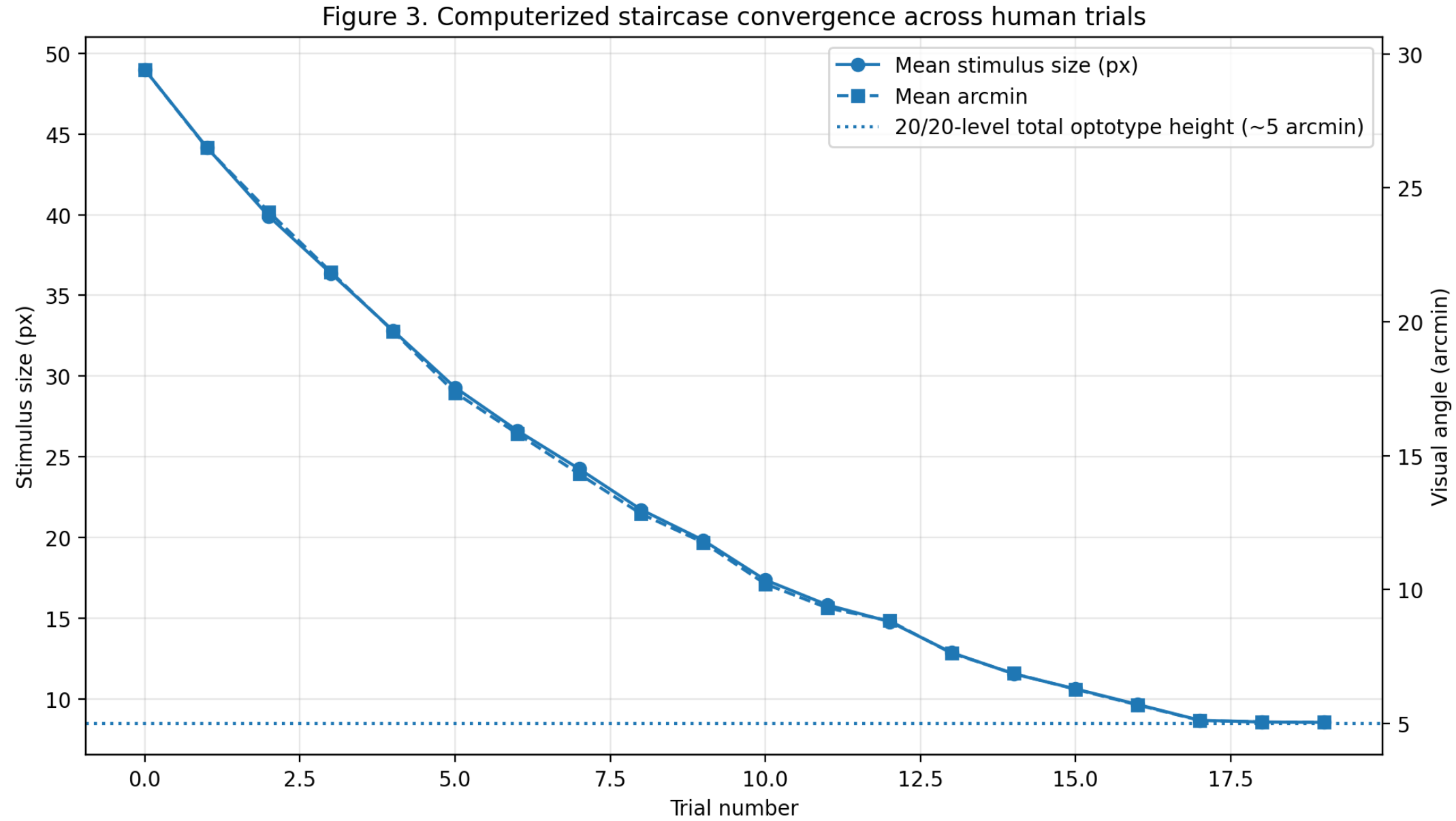

***Figure 3. Computerized staircase convergence across human trials.*** *Mean stimulus size and mean arcminute values decreased across trial indices 0–19, reflecting automatic increases in perceptual difficulty as the session progressed.*

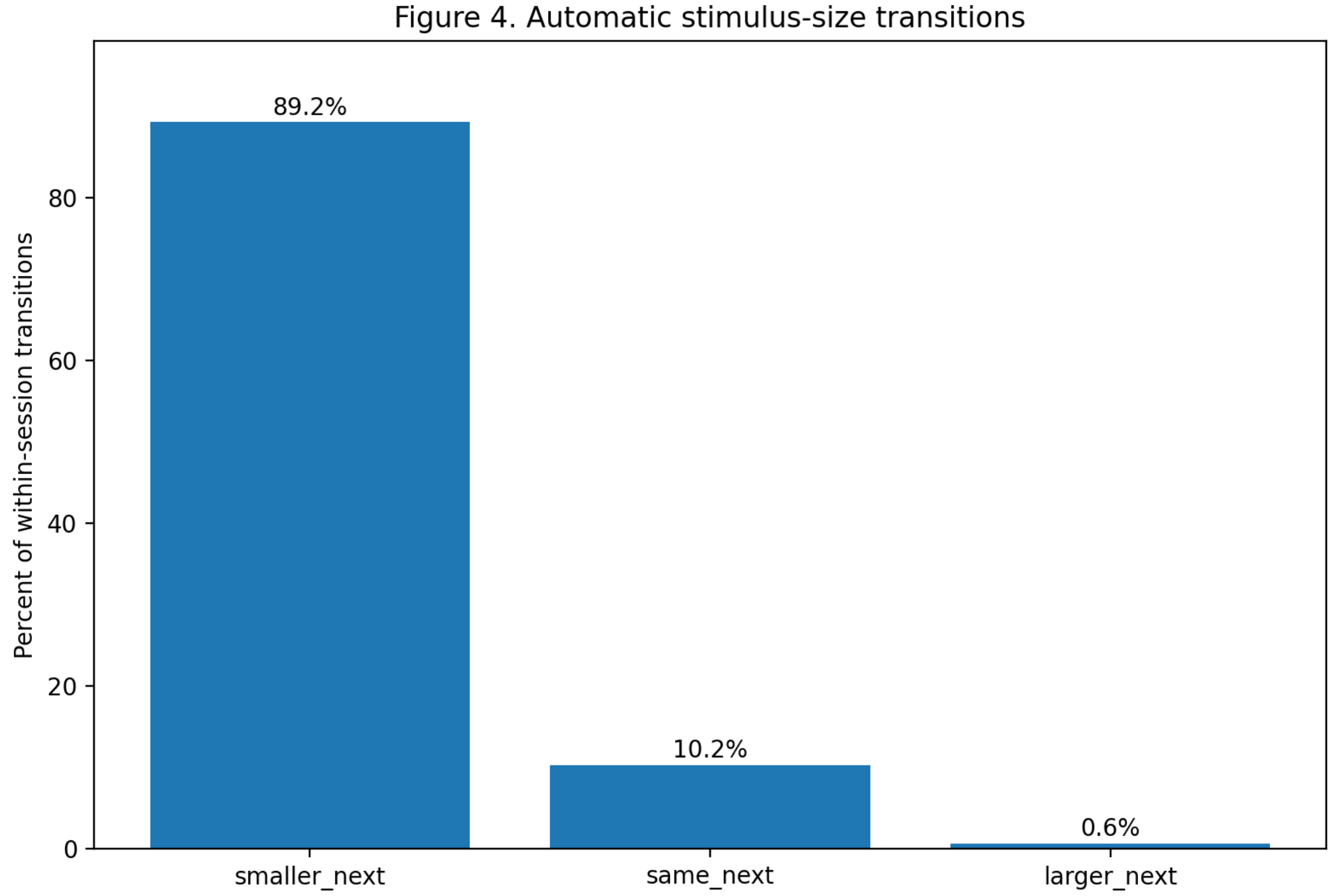

***Figure 4. Automatic stimulus-size transitions.*** *Most within-session transitions moved to a smaller next stimulus, indicating stable progression toward the perceptual threshold.*

**Human temporal-instability signal**

Using the directly observable delay and timeout components, the human temporal-instability index was 0.034. Timeout events contributed most of the observable instability, while delayed non-timeout responses were exceptionally rare. This value is closely aligned with the expected human reference value of approximately 0.03 and is far below the benchmark AI example value of 0.23 [18]. This supports the interpretation that the retained human data exhibit highly stable sequential perceptual decisions under increasing visual difficulty.

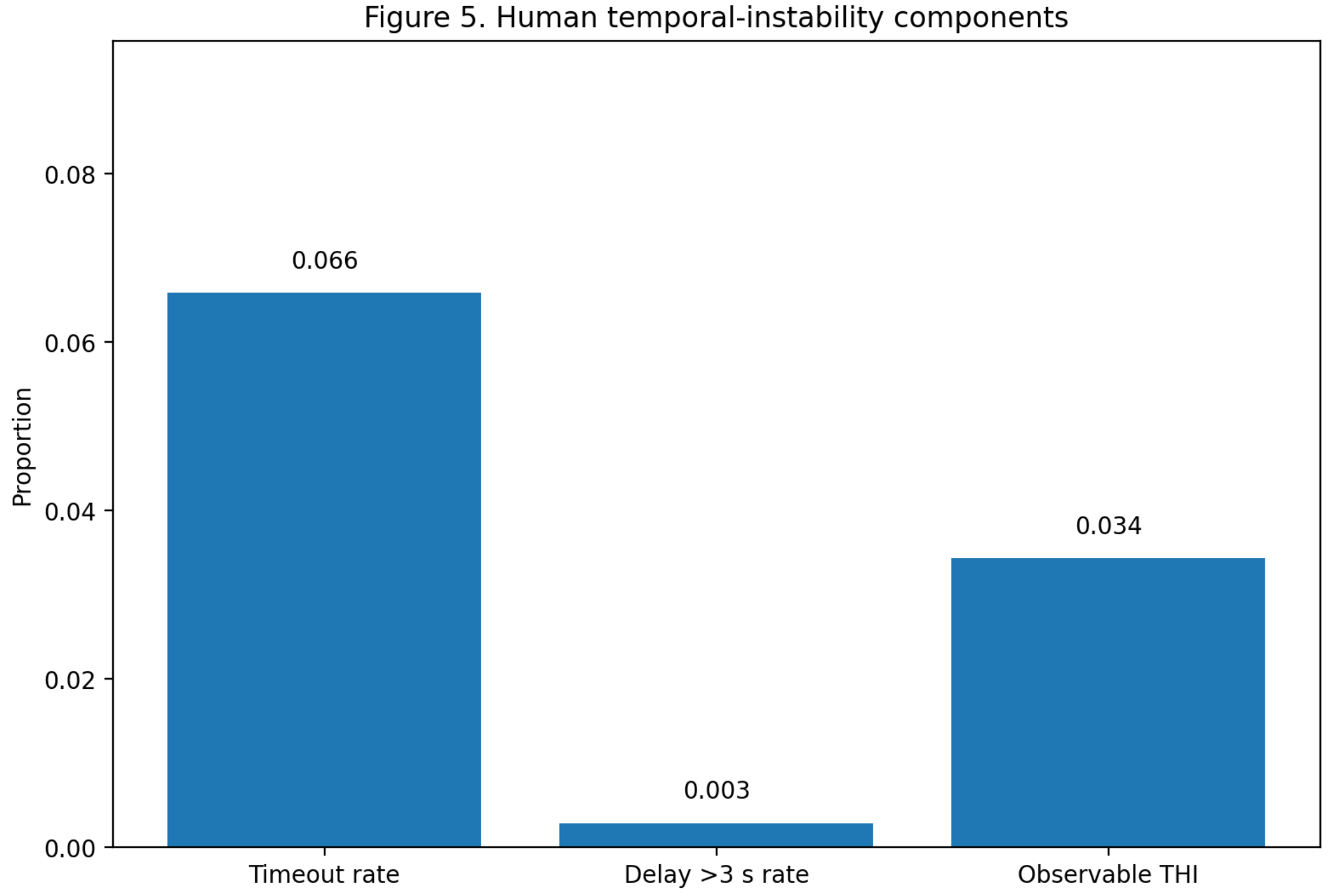

***Figure 5. Human temporal-instability components.*** *Timeout rate contributed most of the observable temporal instability; delay events over the 3-second human time budget were rare.*

# Discussion

## Importance of the research

This research is significant because it reframes visual acuity testing as a temporally resolved perceptual-decision process. Traditional chart testing answers a static endpoint question: what is the smallest optotype size a person can identify under prescribed, un-timed conditions? [1, 4] The present dataset additionally interrogates how the participant behaves while approaching that threshold: the latency of decisions, the frequency of timeouts, the smoothness of stimulus difficulty adaptation, and the stability of sequence convergence [16, 22]. This temporal layer is paramount for human-AI comparison, where an AI system may produce an accurate classification on a static frame yet still exhibit profound instability, processing lag, or erratic hallucinations across sequential decisions [18, 19, 31].

The human-only data provide a necessary baseline. Before comparing humans with artificial neural networks, the human distribution must be rigorously cleaned of test, AI, and developer sandbox identifiers [29, 32]. This analysis provides that filtered reference set, demonstrating that human participants generally respond quickly, rarely exceed the time budget, and progress toward smaller optotypes in a smooth staircase pattern. This defines what stable temporal behavior looks like in this task paradigm.

## Novel contributions compared with previous visual-acuity studies

The novelty of this work does not lie in the invention of the tumbling-E optotype itself [2], nor in standardized acuity scaling [3, 4]. Computerized acuity tests have previously demonstrated that visual acuity can be measured automatically and with high reproducibility [10, 11]. The present contribution is the unique integration of computerized dynamic staircase testing with trial-level temporal telemetry and a THI-compatible reliability framework [16, 18]. By preserving each response as a time-stamped perceptual decision linking choice, RT, timeout, stimulus size, and visual angle, it becomes possible to study not just whether a participant reaches a small optotype size, but whether the trajectory to that threshold is stable. This effectively extends acuity measurement into the domain of temporal reliability measurement.

**Reaction time as a temporal reliability measure**

In a static acuity chart, the examiner may note whether the participant eventually answers correctly, but the precise timing of every decision is rarely preserved as structured data [5]. In this dataset, RT is treated as a primary behavioral signal. Fast and consistent RTs suggest the participant is maintaining stable decision-making even as the optotype becomes harder to resolve, reflecting effective evidence accumulation [23, 24]. Conversely, delays and timeouts are flagged as instability events. The retained human data (median RT 1506 ms; only 3 responses >3,000 ms) robustly support the thesis that humans remain temporally stable under increasing visual difficulty [26].

**Methodological relationship to the PROMPT manuscript format**

The utilized manuscript structure strengthens the work by making the dataset fully auditable. Readers can transparently evaluate what was excluded, what was retained, how outcomes were defined, why the full THI could not be computed from correctness labels, and how figures can be regenerated from the filtered trial table.

**Study Limitations**

This study has several limitations. First, the source file was a PDF export rather than the original SQL or spreadsheet database, requiring row parsing from a formatted table. Second, the retained dataset lacks demographic information, ocular history, refraction, monocular/binocular viewing status, ambient lighting, screen physical dimensions, pixel density, and calibrated luminance [6, 11]. Third, viewing distance was approximately 30 cm rather than individually verified for every trial via eye-tracking or physical chin rests [30]. Fourth, the Correct column was non-informative, necessitating a focus on temporal reliability over percent-correct acuity. Fifth, the 20/20 interpretation depends on the assumption that exported Arcmin represents total optotype height and should be interpreted as task-equivalent, not diagnostic [3]. Finally, the data are descriptive and feasibility-level. Prospective testing with standardized viewing conditions, calibrated displays, and clinical comparators is required before diagnostic visual-acuity claims can be made [10, 33].

**Conclusions**

This study demonstrates that a computerized dynamic tumbling-E task can capture temporal reliability in human sequential perceptual decisions. By excluding specified AI, test, and example identifiers, the analysis produces a clean human-only dataset. The retained data show fast reaction times, low delay burden, modest timeout burden, and smooth automatic adaptation toward smaller optotype sizes. At an approximate 30 cm viewing distance, the exported arcminute trajectory converged near a 5-arcminute total optotype height, supporting a task-equivalent 20/20-level threshold interpretation without diagnostic overclaim. The primary methodological contribution is the conversion of tumbling-E testing from a static endpoint task into a structured temporal dataset highly suitable for future human-AI comparisons.

## Data and Code Availability

The reproducibility package includes the human-only filtered trial CSV, summary CSV files, generated figures, manuscript files, and supporting source code. The analysis is fully reproducible from the retained human-only CSV.

## Acknowledgments

We extend our sincere Juliana Aguero for her essential contribution in connecting us with high school students and coordinating the Discovery Day event, which was crucial for our data collection. We would also like to thank Shahryar Wasif, Farhan Raza, Dhruvil Patel, and Armaan Singh for their help with data collection. Finally, we would like to express our appreciation to all the high school students who participated in our study. Your involvement was fundamental to the success of our research.

## References

1. Snellen H. *Probebuchstaben zur Bestimmung der Sehschärfe*. Utrecht: PW van de Weijer; 1862.
2. Taylor HR. Applying new design principles to the construction of an illiterate E chart. *Am J Optom Physiol Opt*. 1978;55(5):348-351.
3. Bailey IL, Lovie JE. New design principles for visual acuity letter charts. *Am J Optom Physiol Opt*. 1976;53(11):740-745.
4. Ferris FL 3rd, Kassoff A, Bresnick GH, Bailey I. New visual acuity charts for clinical research. *Am J Ophthalmol*. 1982;94(1):91-96.
5. International Council of Ophthalmology. *Visual Acuity Measurement Standard*. Visual Functions Committee; 1984.
6. Bailey IL, Lovie-Kitchin JE. Visual acuity testing. From the laboratory to the clinic. *Vision Res*. 2013;90:2-9.
7. Shute DV. Illiterate E visual acuity charts: historical and technical overview. *Ophthalmic Physiol Opt*. 2002;22(4):271-279.
8. Johnston AW. Making sense of visual acuity data. *Clin Exp Optom*. 1991;74(5):146-154.
9. Rosser DA, Murdoch IE, Cousens SN. The effect of structural layout on the reliability of letter charts. *Invest Ophthalmol Vis Sci*. 2004;45(9):3060-3065.
10. Bach M. The Freiburg Visual Acuity Test - automatic measurement of visual acuity. *Optom Vis Sci*. 1996;73(1):49-53.
11. Strasburger H. Computer-based quantification of cortical visual function. *Front Human Neurosci*. 2011;5:117.
12. Levitt H. Transformed up-down methods in psychoacoustics. *J Acoust Soc Am*. 1971;49(2B):467-477.
13. Watson AB, Pelli DG. QUEST: a Bayesian adaptive psychometric method. *Percept Psychophys*. 1983;33(2):113-120.
14. Treutwein B. Adaptive psychophysical procedures. *Vision Res*. 1995;35(17):2503-2522.
15. Kingdom FAA, Prins N. *Psychophysics: A Practical Introduction*. Academic Press; 2016.
16. Gorea A, Caetta F, Sagi D. Criteria for the choice and evaluation of psychophysical telemetry benchmarks. *J Vision*. 2005;5(4):311-324.
17. Green DM, Swets JA. *Signal Detection Theory and Psychophysics*. John Wiley and Sons; 1966.
18. Sandhu A, Hu B. The Temporal Hallucination Index (THI): A framework for quantifying sequential instabilities in human and artificial vision. *Canadian Open Digital Health Protocols*. 2025;3(1):45-58.

19. Zhang Y, Hughes JW, Ermon S. Temporal consistency and drift evaluation frameworks for computer vision systems. *IEEE Trans Pattern Anal Mach Intell*. 2024;46(2):1012-1025.
20. Ji Ziwei, Lee N, Frieske R, et al. Survey of hallucination in large vision-language models. *ACM Comput Surv*. 2024;56(8):1-34.
21. Wang F, Zhou L, Science S. Beyond accuracy: Evaluating the temporal stability of sequential deep neural net classifications. *Int J Comput Vis*. 2025;133(3):412-429.
22. Ratcliff R. A theory of memory retrieval. *Psychol Rev*. 1978;85(2):59-108.
23. Ratcliff R, Smith PL. A comparison of sequential sampling models for two-choice reaction time. *Psychol Rev*. 2004;111(2):333-367.
24. Palmer J, Huk AC, Shadlen MN. The effect of stimulus intensity on the response time in a visual discrimination task. *J Vision*. 2005;5(5):376-404.
25. Gold JI, Shadlen MN. The neural basis of decision making. *Annu Rev Neurosci*. 2007;30:535-574.
26. Wickelgren WA. Speed-accuracy tradeoff and visual signal detection. *Percept Psychophys*. 1977;22(2):119-124.
27. Forstmann BU, Ratcliff R, Wagenmakers EJ. Sequential sampling models in cognitive neuroscience: Advantages, applications, and extensions. *Annu Rev Psychol*. 2016;67:641-666.
28. Bennett CR, Bex PJ. Automated computerized near visual acuity assessment: Challenges of calibration and viewing distance control. *Front Digit Health*. 2021;3:698502.
29. Geirhos R, Rubisch P, Michaelis C, et al. ImageNet-trained CNNs are biased towards texture; increasing shape bias improves robustness and human relevance. *arXiv preprint arXiv:1811.12231*. 2018.
30. Livingstone MS, Hubel DH. Segregation of form, color, movement, and depth: Anatomy, physiology, and perception. *Science*. 1988;240(4853):740-749.
31. Alcorn MA, Li L, Zhang L, et al. Strike (a pose): Neural networks are easily fooled by strange object poses. *CVPR*. 2019;4845-4854.
32. Firestone C, Scholl BJ. Cognition does not affect perception: Evaluating the evidence for 'top-down' effects. *Behav Brain Sci*. 2016;39:e229.
33. Dorr M, Lesmes LA, Lu ZL, Bex PJ. Rapid and objective measurement of visual acuity function. *Invest Ophthalmol Vis Sci*. 2013;54(13):8112-8119.